\documentclass[pr,aps,praps,amsbsy,superscriptaddress]{revtex4}
\usepackage{graphics}
\usepackage{color} 

\vspace {7 mm}

\begin{document}

\title{Reduction of multiple scattering of high-energy positively charged particles during channeling in single crystals }

\author{W.~Scandale}
\affiliation{CERN, European Organization for Nuclear Research, CH-1211 Geneva 23, Switzerland}
\author{L.S.~Esposito}
\affiliation{CERN, European Organization for Nuclear Research, CH-1211 Geneva 23, Switzerland}
\author{M.~Garattini}
\affiliation{CERN, European Organization for Nuclear Research, CH-1211 Geneva 23, Switzerland}
\affiliation{ Imperial College, London, United Kingdom}

\author{R.~Rossi}
\affiliation{CERN, European Organization for Nuclear Research, CH-1211 Geneva 23, Switzerland}
\author{V.~Zhovkovska}
\affiliation{CERN, European Organization for Nuclear Research, CH-1211 Geneva 23, Switzerland}
\author{A.~Natochii} 
\affiliation{CERN, European Organization for Nuclear Research, CH-1211 Geneva 23, Switzerland}
\affiliation{Laboratore de l'Accelerateur Lineaire (LAL), Universite Paris Sud Orsay, Orsay, France}


\author{F.~Addesa} 
\affiliation{INFN Sezione di Roma Roma, Piazzale Aldo Moro 2, 00185 Rome, Italy}
\author{F.~Iacoangeli}
\affiliation{INFN Sezione di Roma Roma, Piazzale Aldo Moro 2, 00185 Rome, Italy}
\author{F.~Galluccio}
\affiliation{INFN Sezione di Napoli, Italy}
\author{F.~Murtas} 
\affiliation{CERN, European Organization for Nuclear Research, CH-1211 Geneva 23, Switzerland}
\affiliation{INFN, LNF, Via Fermi, 40 00044 Frascati (Roma) , Italy}
\author{A.G.~Afonin}
\affiliation{NRC Kurchatov Institute - IHEP, 142281, Protvino, Russia} 
\author{Yu.A.~Chesnokov}
\affiliation{NRC Kurchatov Institute - IHEP, 142281, Protvino, Russia} 
\author{A.A.~Durum}
\affiliation{NRC Kurchatov Institute - IHEP, 142281, Protvino, Russia}
\author{V.A.~Maisheev\footnote[1]{Corresponding author: maisheev@ihep.ru} }
\affiliation{NRC Kurchatov Institute - IHEP, 142281, Protvino, Russia} 
\author{Yu.E.~Sandomirskiy}
\affiliation{NRC Kurchatov Institute - IHEP, 142281, Protvino, Russia}
\author{A.A.~Yanovich}
\affiliation{NRC Kurchatov Institute - IHEP, 142281, Protvino, Russia}
\author{G.I.~Smirnov}
\affiliation{CERN, European Organization for Nuclear Research, CH-1211 Geneva 23, Switzerland}
\affiliation{Joint Institute for Nuclear Research, Joliot-Curie 6,141980, Dubna, Russia}  
\author{ Yu.A.~Gavrikov} 
\affiliation{Petersburg Nuclear Physics Institute in National Research Centre "Kurchatov Institute",
 188300, Gatchina, Russia}
\author{Yu.M.~Ivanov}
\affiliation{Petersburg Nuclear Physics Institute in National Research Centre "Kurchatov Institute",
 188300, Gatchina, Russia}
\author{M.A.~Koznov} 
\affiliation{Petersburg Nuclear Physics Institute in National Research Centre "Kurchatov Institute",
 188300, Gatchina, Russia}
\author{M.V.~Malkov}
\affiliation{Petersburg Nuclear Physics Institute in National Research Centre "Kurchatov Institute",
 188300, Gatchina, Russia}
\author{L.G.~Malyarenko}
\affiliation{Petersburg Nuclear Physics Institute in National Research Centre "Kurchatov Institute",
 188300, Gatchina, Russia}
\author{I.G.~Mamunct}
\affiliation{Petersburg Nuclear Physics Institute in National Research Centre "Kurchatov Institute",
 188300, Gatchina, Russia}
\author{J.~Borg}
\affiliation{ Imperial College, London, United Kingdom}
\author{T.~James}
\affiliation{ Imperial College, London, United Kingdom}
\author{G.~Hall}
\affiliation{ Imperial College, London, United Kingdom}
\author{M.~Pesaresi}
\affiliation{ Imperial College, London, United Kingdom}

\begin{abstract}
We present the experimental observation of the reduction of multiple scattering of high-energy positively charged particles during channeling in single crystals. 
 According to our measurements the rms angle of multiple scattering in the  plane orthogonal to the  plane of the
channeling is  less  than half that for non-channeled particles moving in the same crystal. In the experiment we use 
focusing bent single crystals. Such crystals have a variable thickness in the direction of beam propagation. This allows us to measure 
 rms angles of scattering as a function of thickness for channeled and non-channeled particles.   
The behaviour with  thickness of non-channeled particles is in agreement with expectations  whereas the behaviour of channeled particles has
unexpected features. We give a semi-quantitative explanation of the observed effect.
\end{abstract}

\maketitle 
\section{Introduction}
It has long been known that the interaction of charged particles with a single crystal differs in many respects from such interactions with 
an amorphous substance \cite{TM,HU,BKS}. In particular, in single crystals, the effect of planar channeling is observed, when positively charged particles
entering the crystal at small angles with respect to the system of crystallographic planes are captured by these planes
between adjacent layers of atoms \cite{JL}. In the plane orthogonal to the crystallographic planes the trajectory of particles
is determined by the atomic planar potential. In the direction along the crystallographic planes the particle motion is practically independent
of the atomic potential but in principle the particle may  interact with individual  positively charged atomic nuclei. The result of such interactions  is multiple scattering of the particle.    

In this paper we investigate the multiple scattering of particles during their channeling within silicon crystals.
The investigation is based on measurements of interactions of high energy (180 and 400 GeV/c) positively charged particles with 
 focusing bent silicon crystals (with (111) planar orientation ) \cite{WS}.
We compare  these results with those for non-channeled particles.    
We have not seen publications with measurements of multiple scattering when channeling positively charged particles in single crystals.
However,  in Ref. \cite{FA} were presented the  results of scattering for non relativistic (3- 10 MeV) protons channeled in silicon and germanium
single crystals.  
Besides, for negatively charged particles (electrons with energy in the range 3.35-14 GeV),  the electron  
scattering angle was 1.7 times larger than predicted for an amorphous medium \cite{Wis}.

The problems of the scattering of relativistic particles in straight single crystals were considered in \cite{BG,LP,SH}.
In the paper \cite{BG} was presented the theory of scattering of relativistic particles in the straight single crystals.
 Here in particular were found the  coherent and incoherent contributions in the total  cross section of the scattering process.
In Ref.\cite{LP} is shown that if an ultrarelativistic charged particle intersects crystallographic planes at a sufficiently
small angle  (but larger than critical angle of channeling), the mean square angle of its deviation in the plane perpendicular to the  crystallographic plane decreases in comparison with the case of an amorphous material.

It should be noted that the scattering processes of ultrarelativistic particles in bent crystals should be differ
from such processes in straight crystals. Formally, it is easy to understand since in a bent crystal plays an important
 role of the so-called effective potential \cite{TS}. So, due to this, in particular, there is a specific type of scattering, called
volume reflection \cite{TV, MV}. It can be expected that between the scattering processes in straight and bent crystals, there is a certain connection, perhaps a similar to connection between the processes of radiation of light leptons or the photoproduction of
 electron-positron  pairs in such crystals and the established
in papers\cite{MV1,BM}.

The paper is organized as follows.
 First, we give a short description of the
experimental setup and describe the procedure of the data
analysis. In the subsequent sections we present the experimental
results obtained with three crystals in the focusing
and defocusing mode, after which a discussion and conclusions
follow.

\section{Experimental setup}

The experiment was carried out at the H8 beam line of
the CERN SPS using a practically pure 400 GeV/c proton
beam and a 180 GeV/c beam of positive secondary
particles for the measurements. The layout of the experiment
shown in Fig. 1 was similar to that described earlier in
\cite{WS}. A high precision goniometer was used to orient the
crystal planes with the respect to the beam axis with an
accuracy of 2 $\mu$rad. The accuracy of the preliminary crystal
alignment with a laser beam was about 0.1 mrad. Five pairs
of silicon microstrip detectors, two upstream and three
downstream of the crystal deflector, were used to measure
incoming and outgoing angles of particles with an angular
resolution in each arm of about 3 $\mu$rad \cite{MP, GH}. 
The detector spatial
resolution was measured to be about 7 $\mu$m.
The geometric
parameters of the incident beam were measured with the
help of the detector telescope. The width of the beam along
the horizontal and vertical axes was a few millimeters. The
angular divergence of the incident beam in the horizontal
and vertical planes was $\sim 10$ $\mu$rad for the proton beam and
$\sim 20$ $\mu$rad (horizontal) and $\sim 40$ $\mu$rad (vertical) for the
secondary beam of positively charged particles. The average
cycle time of the SPS during the measurements was
about 45 s with a pulse duration of 10-11 s. Typically, the
number of particles per spill was $1.3 \, 10^6$.

The experimental configuration described here was used to measure the focusing properties of special single crystals.
Such crystals were placed in the goniometer and after the orientation procedure the envelope of the beam was measured 
(see Fig. 1). Several focusing single crystals were investigated in focusing or defocusing modes.
Part of the data obtained in experiments with focusing crystals was used for the present study. 

\section{Focusing single crystals}

The first measurements of the beam focusing effect were performed in the 1990s \cite{Fok}.
Since then the focusing devices have been significantly improved \cite{WS,WS2}. Fig. 2 illustrates the
operation principle of such devices. The focusing crystal is represented by the sum of rectangle
ABCF and triangle FCD (see Fig. 2a). Positively charged particles entering the bent
crystal in the channeling regime are deflected through the same angle over the distance BC
(AF). For a sufficiently large deflection angle, the channeled and non-channeled particles
(background) are spatially separated (see also Fig. 1). The triangular part of the crystal 
deflects particles
with different transverse coordinates $x$ according to a linear relationship between the angle
and coordinate. Therefore, the particle trajectories converge at some (focal) point (see Fig. 2b). The
results of a recent study of strip focusing crystals can be found in  \cite{WS}.

The list of  focusing crystals investigated is presented in Table 1 of  \cite{WS}. 
For the present study we use the data collected for crystals 1, 2 and 4.
 Crystal 1 has dimensions: $AB=2.07\pm 0.01$  mm,\, $AD=49.84 \pm 0.02$ mm\, $BC=29.8\,$ mm.
Crystal 2 is approximately the same size.
Crystal 4 has the same $AD$ and $BC$ sizes as  crystal 1 but $AB\approx 4\, $mm. 
The height (not shown in the figure) of every crystal was  $\approx 86\,$mm.

\section{Measurements}

The procedure for spatial and angular orientation of the crystal in the goniometer was described in our previous articles \cite{WS3}.
As a result, for every particle crossing the oriented crystal we obtained: a) the horizontal and vertical coordinates;
b) the horizontal and vertical incident angles; c) the horizontal and vertical outgoing angles after the crystal.  
The difference between the horizontal (vertical) outgoing and incoming angles gives the horizontal (vertical) deflection angle for each particle. 
Fig. 2b illustrates the results in a two-dimensional plot of the horizontal angle of deflection versus the horizontal coordinate. 
The particles captured in the channeling regime and which passed through the body of crystal in this regime 
are located   between lines $Q_+Q'_+$ and $Q_- Q'_-$.   
Selected in this way the set of channeled particles undergoes multiple scattering in the vertical direction.
Distributions of channeled particles over vertical scattered angles constitute one of the subjects of our study.

For comparison, we also need vertical angular distributions of non-channeled particles. Such particles were selected
among particles with an angle of entry into the crystal that exceeds the critical channeling angle. We used  data only for particles
with an angle of entry directed in the opposite direction to the angle of deflection due to bending.

In addition, the data obtained for each case (channeled and non-channeled particles) were divided into 20 parts for 
crystals 1 and 2 (40 parts for the  crystal 4) according to their horizontal coordinates.  So, in section 0 we took  particles with 
 horizontal coordinates from -0.05 to 0.05  mm, in section 1 those with  horizontal coordinates from 0.05 to 0.15 mm,
in section -1 those with  horizontal coordinates from -0.05 to -0.15 mm  and so on. Such a selection of the data 
allowed us to study the process of multiple scattering for different thicknesses.  

\section{Analysis}

Accordingly to theoretical expectation, planar  small angle multiple scattering is described by the Gaussian distribution
\begin{equation}
\rho(\theta)= {dN\over d\theta} = {1\over \sqrt{2\pi} \sigma}  \exp{-{(\theta -\bar{\theta})^2\over 2\sigma^2 } },
\end{equation}
where $\theta$ is the deflection angle relative to the mean angle $\bar{\theta}$. The value $\sigma$ is the rms of the distribution.
We will investigate  three different cases of scattering for which we introduce the following definitions:
$\sigma_c$, $\bar{\theta_c}$, $\sigma_y$, $\bar{\theta_y}$ and  $\sigma_x$, $\bar{\theta_x}$ are rms and mean values for
scattering in the vertical plane during channeling, and in the vertical and horizontal planes far from channeling conditions.

It should be noted that Eq.(1) is valid only for relatively small angles of scattering. At large scattering angles the function $\rho(\theta)$
decreases more slowly than follows from Eq.(1). The Gaussian approximation is sufficient  for the central 98\% of the projected
angular distribution, with a width given by \cite{MS}
\begin{equation}
\sigma_0= {13.6 [MeV] \over \beta c p} \sqrt{l/X_0}[1+0.038\ln(l/X_0)] 
\end{equation}
where $p$ and $\beta c$ are the momentum and velocity of the incident particle, $c$ is the velocity of light, $l, X_0$ are the thickness of the scattering medium and its radiation length.

We obtain from Eq.(1):
\begin{equation}
\ln(\rho(\theta))= \ln({1\over \sqrt{2\pi} \sigma} ) -{(\theta -\bar{\theta})^2\over 2\sigma^2 } 
\end{equation}
We see from this equation that the function $\ln(\rho(\theta))$ is linear in $(\theta -\bar{\theta})^2$. 
Fig. 3 demonstrates  the corresponding typical experimental behaviour for one of the $x$-intervals. 
We see that for not too large values of $(\theta-\bar{\theta})^2 $ the experimental data coincide with a straight line.
The solid lines in this figure were obtained by the method of least squares fitting \cite{Be}. 
The range of the fits was up to $\sim 1500 \, \mu rad^2$ for non-channeled particles and
 up to  $\sim 1000 \, \mu rad^2$ for channeled particles. 

From these approximations
we find the values $\sigma_c,\sigma_y,\sigma_x$.
Fig. 4 shows the experimental distributions of $\rho(\theta)$-functions  fitted to Gaussian profiles and the resulting values of
 $\sigma_c,\sigma_y,\sigma_x$.
In a similar way we found the corresponding rms $\sigma_c,\sigma_y,\sigma_x$ for each interval  for crystal 1, and  
$\sigma_c,\sigma_y$  for crystals 2 and 4. This choice we will discuss below.

The dependence of $\sigma_c,\sigma_y,\sigma_x$ as a function of transverse horizontal $x$ (lower scale) and longitudinal $z$ (upper scale)  
coordinates are shown in Figs. 5 and 6. 
The linear connection between $x$ and $z$-coordinates was
\begin{equation}
z[{\rm mm}] = 40 + x_0[{\rm mm}] + kx[{\rm mm}]
\end{equation} 
where the coefficient $k$ is equal to 9.7  for  crystals 1 and 2, and 5 for  crystal 4. The variable 
$x$ was from $\approx -1 $ to 1 mm for crystals 1 and 2, and from $\approx -2 $ to 2 mm
for crystal 4 ($x_0\approx 0)$. We  also show the dependence of the rms value calculated in accordance with Eq.(2). 
We use 9.363 cm for  the radiation length of silicon  \cite{Be}.

The $\sigma$ measurement results were fitted with a linear function $a+bx$. Of course, the dependence of rms  multiple scattering is 
described by square root functions , but for our case of a relatively short distance ($40\pm 10$ mm) the linear approximation
is  good enough.    

It should be noted that our collaboration (UA9) has carried out measurements of multiple scattering of 400 GeV/c protons \cite{WS4}
in single silicon crystals with orientations far from axial or planar channeling. 
That experiment was performed in parallel and at the same time and on the same installation as the experiments described here with focusing crystals.
In  \cite{WS4} the background conditions were investigated. They showed that there is additional scattering of the protons
on material in the beam (strip detectors and other matter). It was suggested to correct the results of measurements of the rms values by the relation 
$\sigma^2_{cor}={\sigma^2_{ms}- \sigma^2_{bg}}$, where $ \sigma_{bg}\approx 5\, \mu$rad. We used this method for correction 
of our rms measurements. However, for 180 GeV/c particles we take $ \sigma_{bg}\approx 10\, \mu$rad (following from Eq.(2)).
 In all the figures we present the measured rms value  and its corrected value.

It is easy to see that the rms multiple scattering (for the same particle energy) differs significantly  for channeled and non-channeled particles.
For non-channeled particles the rms value is an increasing function of the thickness. 
For channeled particles this value is a weakly changing
function of the thickness. It is interesting that for 400 GeV/c particles the rms is an increasing function of the thickness while
for 180 GeV/c particles the analogous function is a decreasing function of the thickness.  

In addition, we have the following interesting observation.  
We computed the value of the function $\Sigma^2 =\int_{-\theta_1}^{\theta_1} \theta^2 \rho(\theta) d\theta$ as a function of $\theta_1$.
In this relation the function $\rho(\theta)$ is in one case the distribution function over vertical scattered angles
of non-channeled particles and in another case the same function for channeled particles; both functions are normalized to unity.
We denote the two  corresponding $\Sigma^2$ functions  as $\Sigma^2_y$ and $\Sigma^2_c$.    
As $\theta_1$ grows, $\Sigma^2_y$ increases in value. This is easy to understand due to the non-Gaussian tail of the observed
particle distributions.  However, for 400 GeV/c the function $ \Sigma^2_c$ is practically constant and coincident with that 
calculated by the least square fits.

We see from Eq.(2) that the rms multiple scattering $\sigma_0$ of relativistic particles (when $\beta \approx 1$ )
is inversely proportional to their momentum (energy) and the $\sigma_0^2$ value  is  proportional to the thickness $l$ (neglecting the logarithmic term). Taking this into account we present in Fig. 7 the results of our linear approximations of $\sigma^2_y$ and $\sigma^2_c$-values divided
by a factor $\kappa^2$ (where  $ \kappa=13.6$[MeV] $10^{-6} /E_0$[MeV]) as  functions of the $z$-variable. Here $E_0=pc$.  

\section{Discussion}

\subsection{Possible explanation of phenomena}

The experimental data demonstrate the reduction  
of multiple scattering for particles moving in the channeling regime in comparison with such processes for non-channeled particles.
This effect has a simple and clear explanation. In amorphous media relativistic particles move at a range of distances (impact parameters)
relative to atomic centres (nuclei). In the theory of multiple scattering  the minimum and maximum impact parameters are given by:
\begin{equation}
b_{max}= {Z^{-1/3}\hbar^2\over m_e e^2};\quad b_{min}= {2Ze^2 \over pv}. 
\end{equation}   
where $Z$ is the atomic number of the material, $m_e$ and $e$ are the mass and charge of the electron, and $v$ is the velocity of the particle.  
Under these conditions the mean square angular value from multiple scattering is given by
\begin{equation}
\bar{\theta}^2=Nl\int_{b_{min}}^{b_{max}}\theta^2(b)2\pi b db={8\pi Nl Z^2e^4\over p^2v^2}\ln{b_{max}\over b_{min}}  
\end{equation}
Here $N$ is the nuclear density, $b$ is the impact parameter of a particle and $\theta(b) = 2Z e^2/(pbv)$.

A particle captured in the planar channeling regime moves in a space between two adjacent crystallographic planes.
At least some of the channeled particles will not be able to get close to the nuclei and hence the scattering angle
will be less than  in amorphous media. Thus the rms multiple scattering angle depends on the distribution of oscillation
amplitudes of channeled particles \cite{Chi}. From this point of view we can explain the abnormal rms dependence of channeled particles
on thickness. Particles with large oscillation amplitudes dechannel during motion and the mean amplitude is decreased.

Eqs. (5) and (6) are obtained in simplified assumptions and are semi-quantitative in nature and serve only for a better understanding of the process. 
A more correct approach to the problem of multiple scattering of particles can be found, for example, in the articles \cite{RG, WSc}.

\subsection{About multiple scattering of non-channeled particles}
The radiation length in amorphous media is determined by the intensity of bremsstrahlung of electrons    
in such media \cite{MS}.
However, it is well known that radiation processes (such as electron bremsstrahlung and electron-positron photoproduction) 
in single crystals  differ from such processes in  amorphous media. The cross sections these processes represent are 
 the sum of two contributions:
coherent and non-coherent. The coherent contribution depends on orientation 
(relative to a beam) of the crystal and, in principle may be 
practically equal to zero (at orientations  far from a narrow angular range). 
The non-coherent contribution is always present  (at any orientation and at constant temperature) 
from  bremsstrahlung and electron-positron pair production      
in the amorphous behaviour of the corresponding crystal. There is only one difference, 
which is a lesser intensity of point-like processes in crystal media compared with amorphous. 
The single crystal media (far from the coherent contribution) 
also may be characterized by the analogous parameter as radiation length. 
But this parameter (in  most cases) is approximately 10\% larger than in amorphous media. 


The function which describes the behavior of the rms angle with  distance (see Eq.(2)) uses the radiation length for amorphous media.
This then raises the question what value should be used for the radiation length. Should it be as for an amorphous medium? Or is it analogous to a crystalline medium?
The experimental  data (see Fig. 7)  show better agreement when the calculated value of the radiation length for an amorphous 
medium is used in the description of multiple scattering.

Figs.5-7 illustrate the behaviour of rms of vertical angle distributions  as the functions of thickness for non-channeled particles.
We see that these functions are close enough to the rms for an amorphous medium (see Eq.(2)).  Taking into account 
results of measurements we got approximation description  of the experimental data in the follow form
\begin{equation}
\sigma_e(z) = {13.6 [MeV] \over E_0} \sqrt{z/X_0^*},
\end{equation}
or in another (equivalent) form
\begin{equation}
\sigma_e(z)= {A\over E_0} \sqrt{z/X_0}
\end{equation}
where values $X_0^* = 99 \pm 7$ mm and $A=13.25 \pm 0.47 $ MeV were found from experimental data at 400 GeV/c.
Slightly worse conditions were at the momentum of the beam 180 GeV/C and crystal number 4. However, we have got
 $X_0^* = 102 \pm 10$ mm and  $A=13.07 \pm 0.65 $ MeV. 
It should be noted that Eq. (2) was obtained for amorphous media with the accuracy equal to 11 \%. 
In this experiment, we studied the multiple scattering of particles in bent silicon single crystals.
Figure 5-7 shows a strong dependence of the process on the initial direction of the particle with respect to
crystallographic planes. Particles not captured in channeling regime in a bent single crystal
are scattered approximately as in the corresponding amorphous medium. However, according to the paper \cite{BG} in the straight crystal
the rms scattering angle should be slightly less by 9 \% than in an amorphous medium. However, measurement errors and nonaccuracy of Eq.(2) 
do not allow us to confirm this prediction.

\subsection{About multiple scattering of channeled particles}

The results presented in Fig. 7 demonstrate  reduction of multiple scattering rms during particle channeling  by 2-2.3 times 
 compared to the passage through the crystal of non-channeled particles. As described above,  the value
$\Sigma_c$ (at a beam momentum of 400 GeV/c) 
 is practically independent of $\theta_1$ starting from about 45 $\mu$rad (see also Fig. 5). 
We ascribe this to reduction of events with large deflection angles due to lack of ability of particles to approach 
close to nuclei.  The comparisons of  curves for the crystal 1 with the curves for crystals 2 and 4  demonstrate approximately inverse proportionality 
with beam momentum
for channeled particles.

On the other hand we see  lesser differences  between  curves for crystals 2 and 4. These  relate to  
particles of 180 GeV/c momentum. We can explain this difference because  crystal 2 and  crystal 4 have different bending radii.
 It is well-known that the motion of particles in bent crystallographic planes can be described with the help of     
the effective potential \cite{TS}: $U(x)+E_0\beta^2 x /R$, where $x$ is the transverse coordinate and $R$ is the bending radius.
The bending radii are $\approx 37$ m and $\approx  200$ m for  crystals 2 and 4, respectively. 
The corresponding  difference
of maximum potentials is about 0.3 eV.  

\section{Conclusions}
1)
The effect of reduction of multiple scattering of positively charged high-energy particles channeled in single crystals was 
experimentally observed for the first time  with high precision.

2)
The multiple scattering behavior of particles as a function of  crystal thickness was experimentally investigated.

3)   
It was shown that the values of characteristic angles of multiple scattering of channeled and non-channeled particles  
are inversely proportional to their energy.

4)
It was shown that the results of measurements on characteristic angles for multiple scattering of non-channeled particles 
are consistent with the known formula describing this process, while for channeled particles we see an unexpected effect that
the rms scattering angle does not increase significantly for thicknesses from 30 to 50 mm. 

 5)
The  effect discovered is also of great practical importance. For example, in \cite{MC, CM} an application 
using two focusing crystals for focusing beams in two orthogonal planes was proposed. 
It can be expected that the effect of reducing multiple scattering will improve the performance of such a system of lenses.

\section{Acknowledgements}
The IHEP participants of UA9 experiment acknowledge financial support of Russian Science Foundation (grant 17-12-01532). 
The Imperial College group thanks the UK Science and Technology Facilities Council for financial support. 
The INFN authors acknowledge the support of the ERC Ideas Consolidator Grant No. 615089 CRYSBEAM. 


\newpage

\begin{figure*} 

\begin{center}
\scalebox{0.8}
{\includegraphics{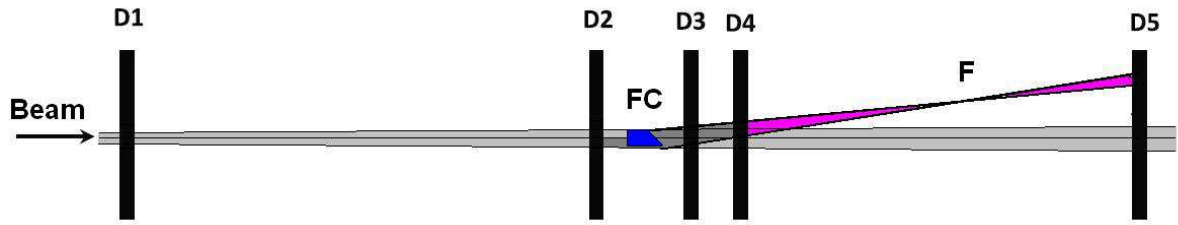}} 
{\caption{
A schematic view of the layout of the experiment for the measurement of multiple scattering of protons in the crystal deflectors. D1-D5 are the silicon microstrip detectors, FC is a bent focusing silicon crystal. Part of the beam after the bent crystal is deflected and focused at point F (focal point).
}}
\end{center}
\end{figure*}



\begin{figure*} 
\begin{center}
\scalebox{0.8}
{\includegraphics{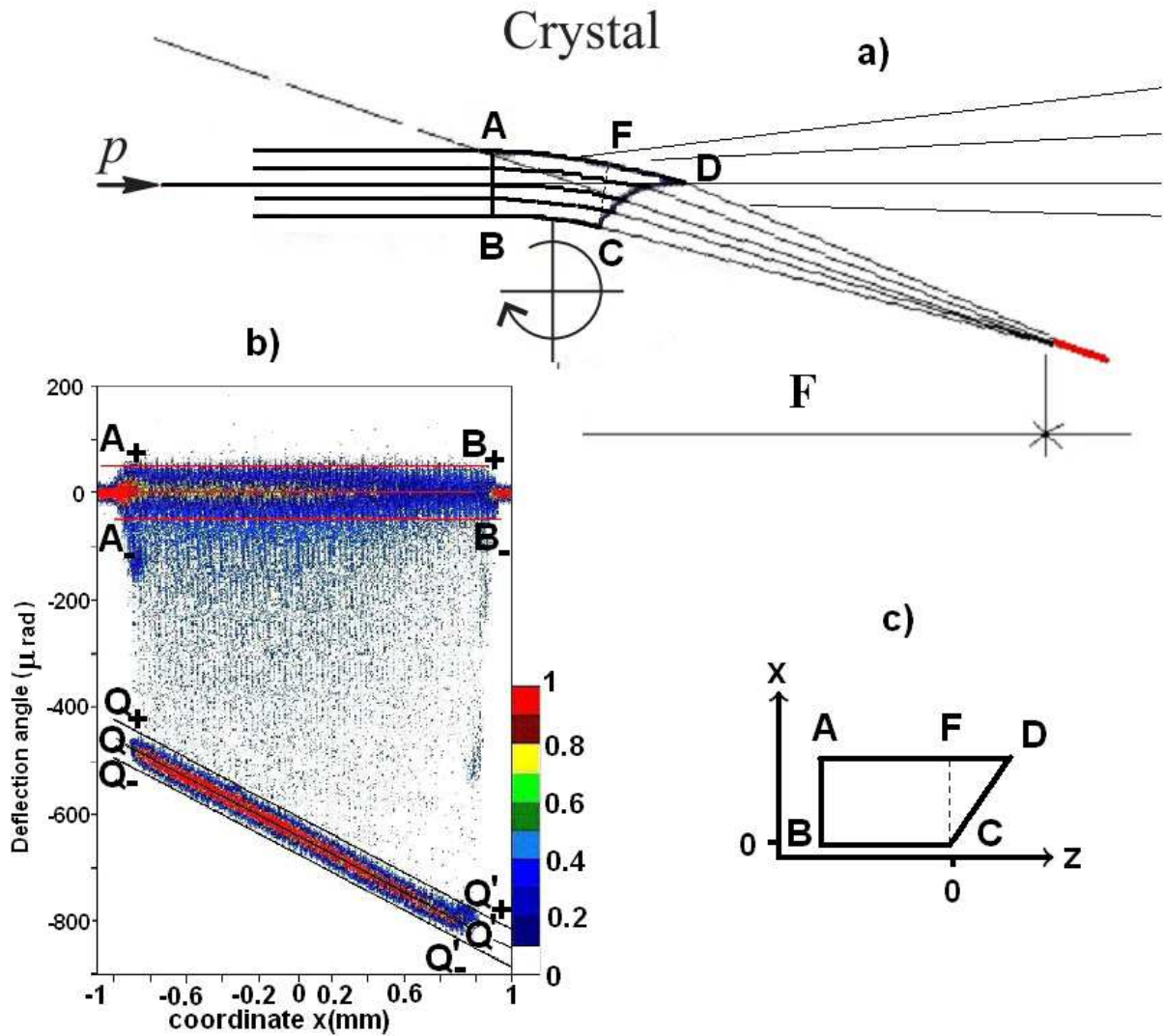}} 
{\caption{ Focusing crystal: a) operation principle; b) the measured two-dimensional plot: deflection angle versus horizontal coordinate $x$;
c) focusing crystal before installation in the holder  
}}
\end{center}
\end{figure*}
 
\begin{figure*} 
\begin{center}
\scalebox{0.8}
{\includegraphics{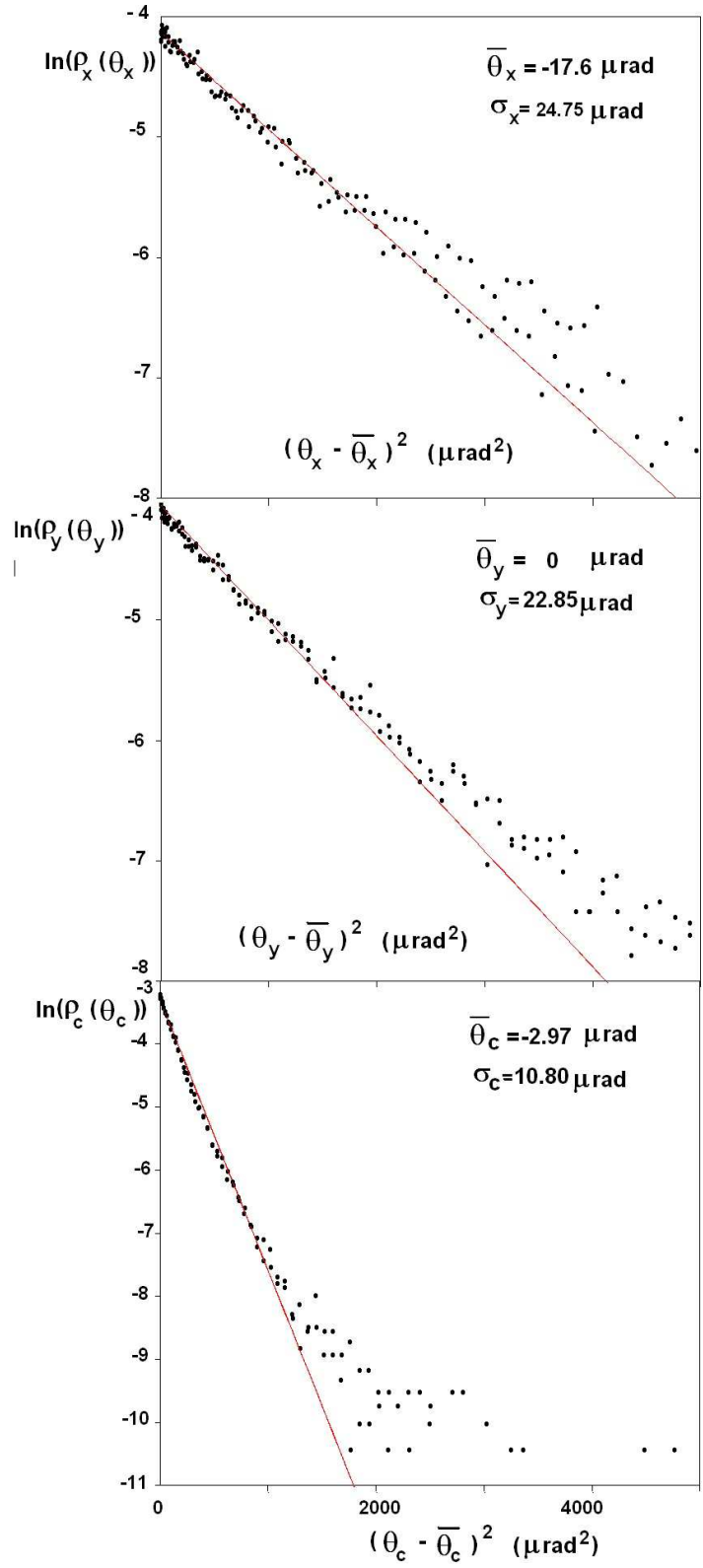}} 
{\caption{ Experimental angular distributions (for horizontal and vertical non-channeled and vertical channeled particles) plotted on a logarithmic scale in accordance with  Eq.(3). The straight lines are fits to linear approximations.
}} 
\end{center}
\end{figure*}

\begin{figure*} 
\begin{center}
\scalebox{0.8}
{\includegraphics{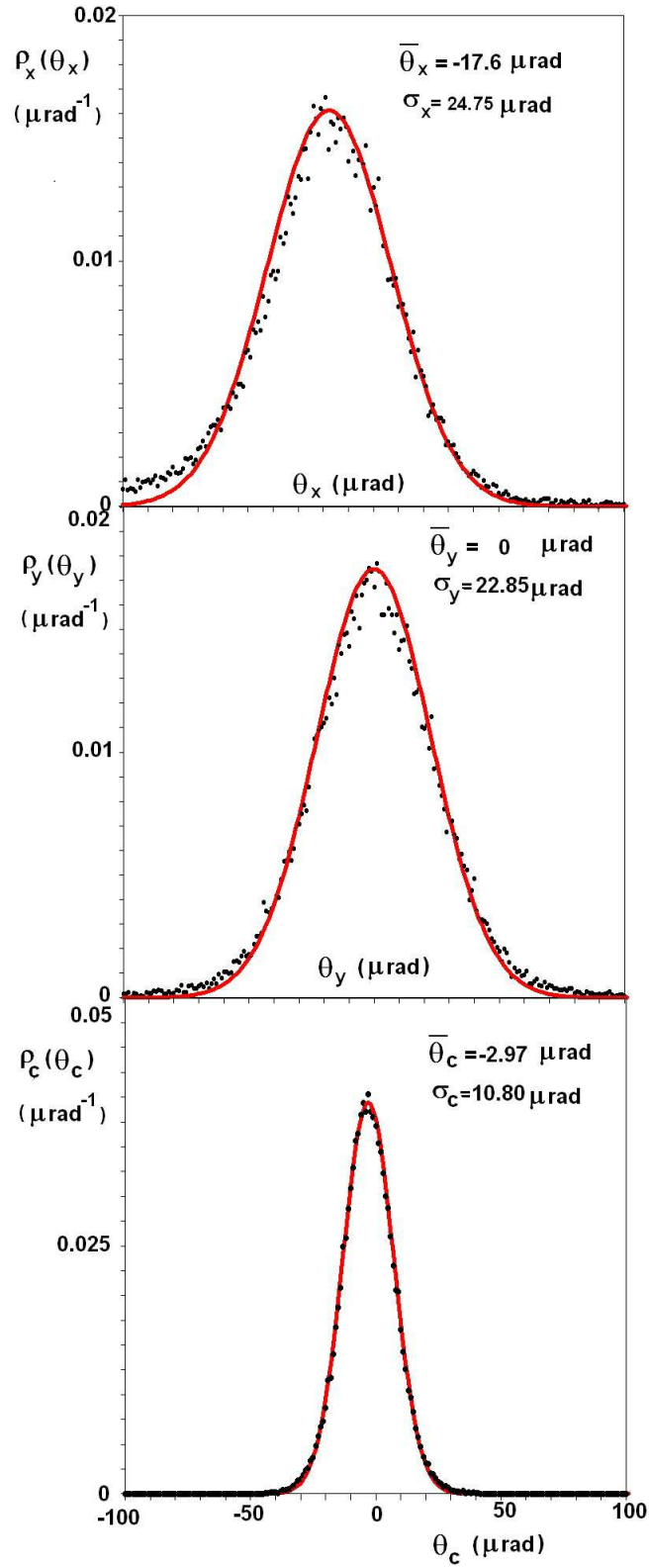}} 
{\caption{ The particle distributions (from Fig. 3) approximated by Gaussian functions.  
}} 
\end{center}
\end{figure*}
\begin{figure*} 
\begin{center}
\scalebox{0.8}
{\includegraphics{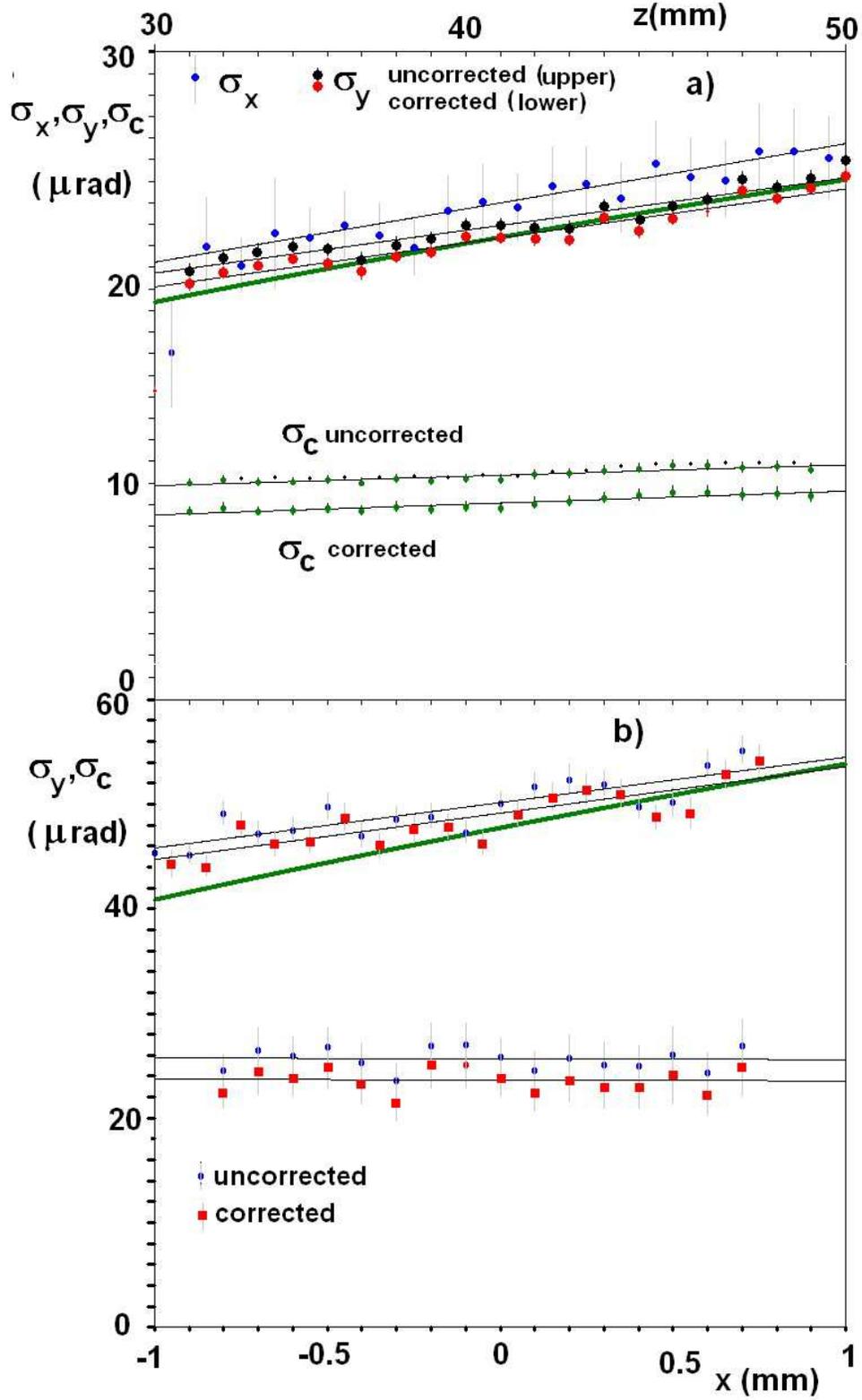}} 
{\caption{The $\sigma_x$, $\sigma_y$ and $\sigma_c$ values as  functions of the distance along the crystal
(from 30 mm up  to 50 mm). The straight lines are  linear approximations 
to the corresponding
experimental points. The data are presented for two cases: a)  beam momentum  400 GeV/c
and  crystal 1; b)  beam momentum 180 GeV/c and  crystal  2.  
The lower horizontal scale is the transverse coordinate $x$ and the upper scale is the corresponding longitudinal coordinate $z$. The meaning of corrected and uncorrected results is explained in the text. 
The small black squares in a)  represent $\Sigma_c$ values ( as determined  in the text).
For better visibility, these squares are offset by 0.05 mm.
The green curves are calculations accordingly to Eq.(2). 
}} 
\end{center}
\end{figure*}
\begin{figure*} 
\begin{center}
\scalebox{0.8}
{\includegraphics{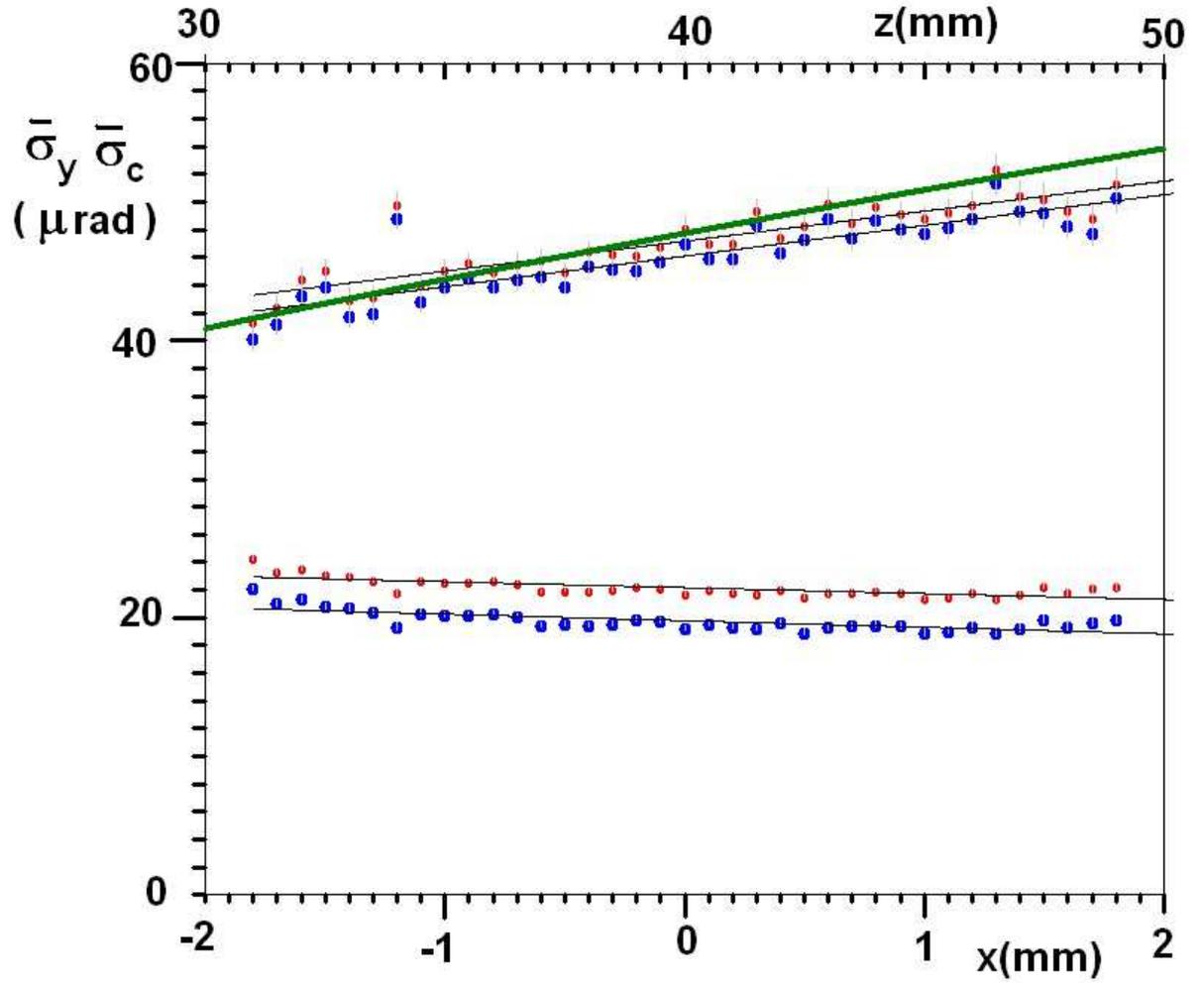}} 
{\caption{The $\sigma_y$ and $\sigma_c$ values as  functions of the distance along 
 crystal  4  (from 30 mm up  to 50 mm). The beam momentum is 180 GeV/c. 
The lower horizontal scale is the transverse coordinate $x$ and the upper scale is the corresponding longitudinal coordinate $z$.
 The red and blue points correspond to uncorrected and corrected (due to backgrownd ) results.
The upper points correspond to $\sigma_y$-values and lower points correspond to $\sigma_c$-values.
The green curves are calculations accordingly to Eq.(2).
}} 
\end{center}
\end{figure*}

\begin{figure*} 
\begin{center}
\scalebox{0.8}
{\includegraphics{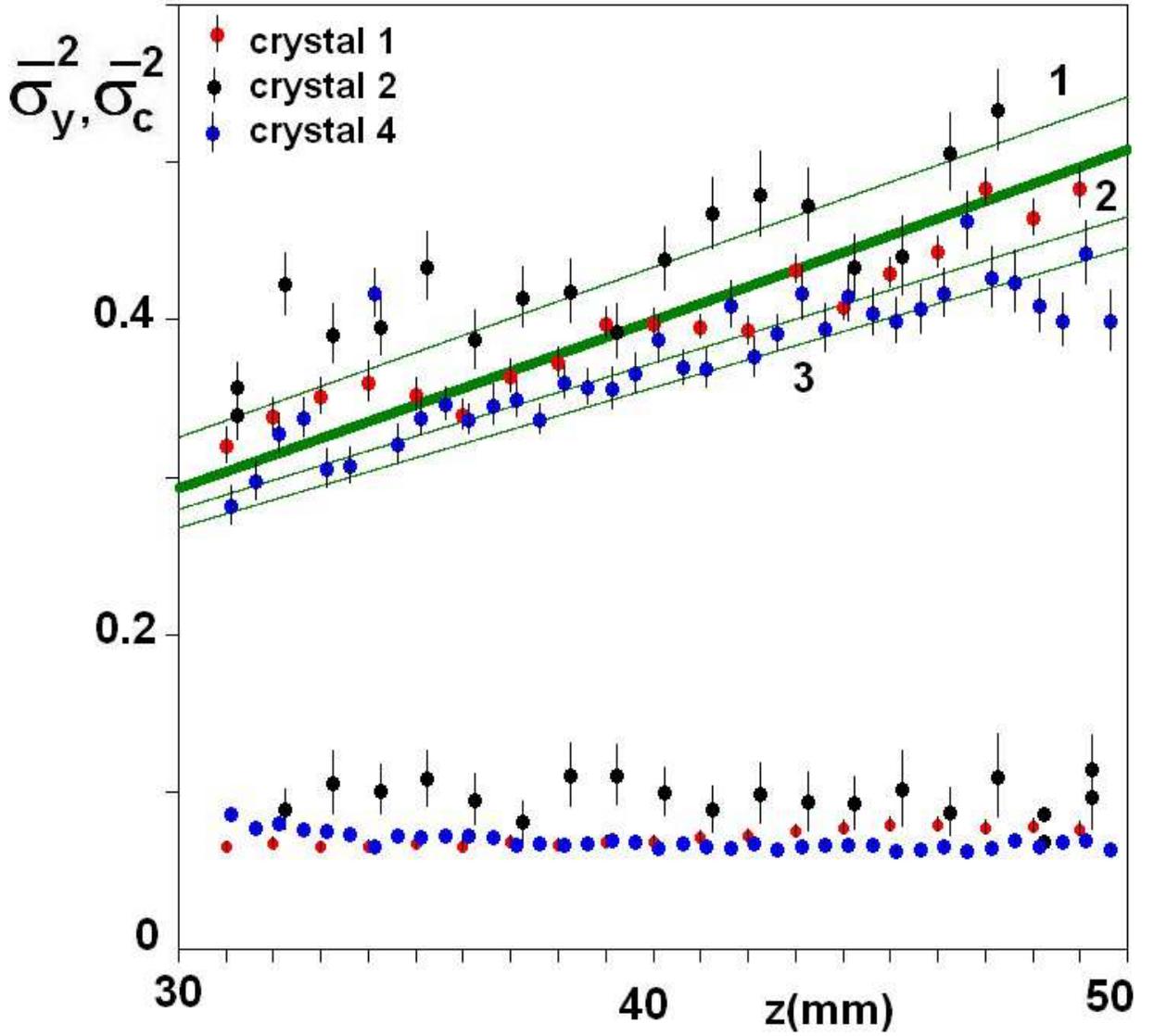}} 
{\caption{The mean square values $ \bar{\sigma_y}^2=\kappa^2 \sigma_y^2, \bar{\sigma_c}^2=\kappa^2 \sigma_c^2$
(where coefficient $\kappa= E_0[{\rm MeV}] \,10^6  /13.6$)  as dependences of the crystal thickness.
The solid thick curve  represents the value $\kappa^2 \sigma_0^2$ (see also Eq.(2)). The dependences were obtained
for all the crystals at condition with the background corrections. The lines 1, 2 and 3 are drawn with accordance of Eqs. (7) and (8)
at boundary values of parameters $A= 13.72, 12.78, 12.42$ MeV and $X_0^* = 92, 106, 112$ mm, correspondingly.  
   }} 
\end{center}
\end{figure*}

\end{document}